\documentclass[aps,pra,floatfix,twocolumn,superscriptaddress,showpacs]{revtex4}
\usepackage{times}
\usepackage{amsmath}
\usepackage{bm}
\usepackage{dcolumn}                 
\newcolumntype{w}[1]{D{.}{.}{#1}}
\newcolumntype{.}{D{x}{}{-1}}

\begin{document}
\title{Electromagnetic moments of the bound system of charged particles}

\author{Albert Wienczek}
\affiliation{
Faculty of Physics, University of Warsaw,
Ho\.{z}a 69, 00-681 Warsaw, Poland} 

\author{Mariusz Puchalski}
\affiliation{Faculty of Chemistry, Adam Mickiewicz University,
             Grunwaldzka 6, 60-780 Pozna{\'n}, Poland}

\author{Krzysztof Pachucki}
\affiliation{
Faculty of Physics, University of Warsaw,
Ho\.{z}a 69, 00-681 Warsaw, Poland} 

\begin{abstract}
We consider a bound system of particles interacting via electromagnetic forces 
in an external electromagnetic field, including leading relativistic corrections. 
Each particle has a definite mass, charge, spin, and charge radius. 
We introduce suitable canonical transformations and a perturbation scheme
to obtain a Hamiltonian which describes the global dynamics of the system.
This enables the derivation of formulas for various electromagnetic moments, 
such as the magnetic dipole moment, the mean square charge radius, and 
the electric dipole polarizability.
\end{abstract}
\maketitle
\section{Introduction}
Let us consider a system of particles, like an atom, an ion, a molecule, or a nucleus,
which forms a bound state. We are interested in the motion and global properties 
of this system in an external electromagnetic field. When relativistic corrections 
are included, the center of mass cannot be separated from  internal degrees of freedom
and this may cause appearance of additional corrections to electromagnetic moments.
A typical problem is the magnetic moment of the system, for which the first correct
description was presented by Hegstrom in \cite{hegstrom}.
Here we aim to present an approach, on the basis of previously obtained results,
which allows one to obtain various electromagnetic moments
including relativistic corrections, e.g. the charge radius and the electric dipole polarizability.
Although we consider electromagnetic systems herein, the approach can be extended
to a nonrelativistic system of strongly interacting particles, such as nuclei.
 
\section{Perturbative approach to separation of the center of mass motion}
We assume that the Hamiltonian $H$ for a system of particles can be decomposed as
\begin{equation}
H =  H_{\rm S} + H_{\Pi} + \delta H, \label{01}
\end{equation} 
where $H_{\rm S}$ is the Hamiltonian that involves only internal degrees of freedom.
$H_{\Pi}$ is the Hamiltonian for the global dynamics of the system, 
which involves center of mass and electromagnetic moments.  
$\delta H$ is the remainder, which couples internal degrees of freedom
to global motion. We will assume that the binding energy is much larger than
the characteristic scale of energy in $\delta H$, so it makes sense to speak about a bound system.
We will also assume that global motion is nonrelativistic. 
We aim to find an effective equation for the global motion that accounts for the coupling to internal degrees of freedom.
The Schr\"odinger equation for the total system is
\begin{equation}
i \frac{\partial \psi}{\partial t} = (H-\mathcal E_{\rm S})\,\psi, \label{02}
\end{equation}
where the leading factor that comes
from the binding energy $\mathcal E_{\rm S}$ was subtracted from the time dependence. 
The Hamiltonians $H_{\rm S}$ and $H_\Pi$ 
involve different degrees of freedom of the system, so they commute
\begin{equation}
\left [ H_{\rm S}, H_\Pi \right ] = 0, \label{03}
\end{equation}
and it allows one to decompose the global wave function as follows
\begin{equation}
\psi = \psi_{\rm S} \psi_\Pi + \delta \psi, \label{04}
\end{equation}
where $\psi_{\rm S}$ is the wave function of the ground state of internal Hamiltonian $H_{\rm S}$ 
with corresponding energy $\mathcal E_{\rm S}$, namely
\begin{equation}
H_{\rm S} \psi_{\rm S} = \mathcal E_{\rm S}\psi_{\rm S}. \label{05}
\end{equation}
$\psi_\Pi$ in Eq. (\ref{04}) is the wave function describing the global dynamics of the system, 
so it depends only on global degrees of freedom. 
The last term in Eq. (\ref{04}), $\delta \psi$, is a small correction that depends on all 
the variables, and results from the coupling of internal and external degrees of freedom.   
We will assume that
\begin{eqnarray}
\langle \psi |\psi \rangle &=& 1 = \langle \psi_{\rm S} | \psi_{\rm S} \rangle_{\rm S},
\label{06}\\
\langle \psi_{\rm S} | \delta \psi \rangle_{\rm S} &=& 0, \label{07}\\
\langle \psi_{\rm S} | \delta H | \psi_{\rm S} \rangle_{\rm S} &=& 0, \label{08} 
\end{eqnarray}
where $\langle\ldots\rangle_S$ denotes the scalar product on internal degrees of freedom only.
If the last condition (\ref{08}) is not satisfied, one can always redefine $H_\Pi$ to include
$\langle \psi_{\rm S} | \delta H | \psi_{\rm S} \rangle_S$, and subtract this expectation
value from $\delta H$. As a result, this condition does not reduce generality.
Let us now project the Schr\"odinger equation (\ref{02}) into $\psi_{\rm S}$, then
using assumptions in (\ref{06}) - (\ref{08}) one obtains
\begin{equation}
i \frac{\partial \psi_{\Pi}}{\partial t} = 
H_{\Pi} \psi_{\Pi} + \langle \psi_{\rm S} | \delta H | \delta \psi \rangle_{\rm S}. \label{09}
\end{equation} 
Eq. (\ref{02}) with the wave function (\ref{04}) can be rewritten in the form
\begin{eqnarray}
\Bigl(\mathcal E_{\rm S} - H_{\rm S} - H_\Pi + i \frac{\partial }{\partial t}\Bigr) \delta \psi &=& 
\psi_S\,\Bigl( H_\Pi - i \frac{\partial }{\partial t}\Bigr)\,\psi_\Pi\nonumber \\&& +
\delta H\,\left ( \psi_{\rm S} \psi_\Pi + \delta \psi \right ) \label{10}
\end{eqnarray}
and  formally
\begin{equation}
 \delta \psi =  \frac{1 }{[(\mathcal E_{\rm S} - H_{\rm S})' -H_\Pi + i\,\partial_t]} 
\delta H \bigl(\psi_{\rm S} \psi_\Pi + \delta\psi\bigr), \label{11}
\end{equation}
where prime denotes exclusion of the ground state from the resolvent, 
which is defined by the following series:
\begin{equation}
 \delta \psi =  \left[
\frac{1}{(\mathcal E_{\rm S} - H_{\rm S})'} + 
\frac{H_\Pi - i\,\partial_t}{(\mathcal E_{\rm S} - H_{\rm S})'^2}+\ldots\right]\,
\delta H \bigl(\psi_{\rm S} \psi_\Pi + \delta\psi\bigr). \label{12}
\end{equation}
$\delta H$ is assumed to be a small correction, 
so we neglect $\delta\psi$ on the right hand side of Eq. (\ref{12})
\begin{equation}
 \delta \psi =  \frac{1 }{[(\mathcal E_{\rm S} - H_{\rm S})' -H_\Pi + i\,\partial_t]} 
\delta H\,\psi_{\rm S} \psi_\Pi. \label{13}
\end{equation}
$H_\Pi$ and the characteristic time scale of $\delta H$ is much smaller
then the excitation energy of $H_S$, so we use Eq. (\ref{12})
and neglect the higher order terms
\begin{eqnarray}
 \delta \psi &=&  \frac{1 }{(\mathcal E_{\rm S} - H_{\rm S})'} \delta H\,\psi_{\rm S} \psi_\Pi
+\ldots \label{14}
\end{eqnarray}
Finally, the equation for $\psi_\Pi$ becomes
\begin{eqnarray}
i \frac{\partial \psi_\Pi}{\partial t} &=& H_{\rm eff}\,\psi_\Pi, \label{15}\\
H_{\rm eff} &=& H_\Pi +
\langle \psi_{\rm S} | \delta H \frac{1}{(\mathcal E_{\rm S} - H_{\rm S})'} \delta H | \psi_{\rm S} \rangle_S
+\ldots \nonumber
\end{eqnarray} 
which can be rewriten in the more convenient form
\begin{equation}
H_{\rm eff} = \langle\psi_S|H-{\cal E}_S|\psi_S\rangle_S 
+ \langle \psi_{\rm S} | H \frac{1}{(\mathcal E_{\rm S} - H_{\rm S})'} H|\psi_{\rm S} \rangle_S
+\ldots \label{15b}
\end{equation}
In the actual calculations we perform additional canonical transformations
to avoid the higher order terms denoted by dots in the above.

\section{Hamiltonian for the compound system}
We consider now a system of $N$ charged particles placed in the external
electromagnetic field, including the leading relativistic corrections.
We assume that the magnetic field is homogenous, and for the electric field
we keep the first derivatives to account for the charge radius.
We would like to separate the center-of-mass motion and obtain general
formulae for electromagnetic moments, such as the magnetic dipole moment $\mu$,
the charge radius, and the electric dipole polarizabiliy. 
Our approach is based on Refs. \cite{compound, magnetic},
where relativistic effects are included perturbatively, while 
the wave function is nonrelativistic and includes the spin.  
The initial Hamiltonian is a sum of  one-particle terms $H_a$ and
two-particle interactions $H_{ab}$ including relativistic corrections
\cite{bethe, lwqed} (using natural units $\hbar=c=1$)
\begin{equation}
H = \sum_a H_a + \sum_{a>b,b} H_{ab}\,, \label{16}
\end{equation}
with
\begin{widetext}
\begin{eqnarray}
H_a &=& \frac{\vec\pi^2_a}{2\,m_a} + e_a\,A^0_a
-\frac{e_a}{2\,m_a}\,g_a\,\vec s_a\cdot\vec B_a
-\frac{e_a}{4\,m_a^2}\,(g_a-1)\,\vec s_a\cdot
\bigl(\vec E_a\times\vec\pi_a-\vec\pi_a\times\vec E_a\bigr)
\nonumber \\ && 
-\frac{\vec\pi^4_a}{8\,m_a^3}
-\frac{e_a}{6}\,r_{Ea}^2\,\nabla\cdot\vec E_a
+\frac{e_a}{8\,m^3_a}\,\Bigl[
4\,\vec\pi_a^2 \, \vec s_a\cdot\vec B_a
+(g_a-2)\,\bigl\{\vec\pi_a\cdot\vec B_a\,,\,\vec\pi_a\cdot\vec s_a\bigr\}\Bigr]
\,, \label{17}\\
H_{ab} &=& \frac{e_a\,e_b}{4\,\pi}\,\biggl\{
\frac{1}{r_{ab}}-\frac{1}{2\,m_a\,m_b}\,\pi_a^i
\biggl(\frac{\delta^{ij}}{r_{ab}}+\frac{r^i_{ab}\,r^j_{ab}}{r^3_{ab}}\biggr)\,
\pi^j_b -\frac{2\,\pi}{3}\,(r_{Ea}^2 +r_{Eb}^2)\,\delta^{3}(r_{ab})
\nonumber \\ &&
+\frac{1}{2\,r^3_{ab}}\,\biggl[
\frac{g_a}{m_a\,m_b}\,\vec s_a\cdot\vec r_{ab}\times\vec \pi_b
-\frac{g_b}{m_a\,m_b}\,\vec s_b\cdot\vec r_{ab}\times\vec \pi_a
+\frac{(g_b-1)}{m_b^2}\,\vec s_b\cdot\vec r_{ab}\times\vec\pi_b
 \label{18}\\ &&
-\frac{(g_a-1)}{m_a^2}\,\vec s_a\cdot\vec r_{ab}\times\vec\pi_a
\biggr]
-\frac{2\,\pi\,g_a\,g_b}{3\,m_a\,m_b}\,\delta^3(r_{ab})\,\vec s_a\cdot\vec s_b
+\frac{g_a\,g_b}{4\,m_a\,m_b}\,\frac{s_a^i\,s_b^j}{r^3_{ab}}\,
\biggl(\delta^{ij}-\frac{3\,r_{ab}^i\,r_{ab}^j}{r_{ab}^2}\biggr)
\biggr\}\,, \nonumber
\end{eqnarray}
\end{widetext}
where $\vec \pi = \vec p-e\,A(\vec r)$. $r_{Ea}^2$ is the mean square charge radius of a
particle $a$ and it includes for convenience the Darwin term, 
so for the point $s=1/2$ particle $r_E^2 = 3/(4\,m^2)$.
We now introduce global variables, the center of mass $\vec R$, 
and the total momentum $\vec \Pi$
\begin{eqnarray}
\vec R &=& \sum_a \frac{m_a}{M}\,\vec r_a\,, \label{19}\\
\vec \Pi &=& \sum_a \bigl[\vec p_a-e_a\,\vec A(\vec R)\bigr] = \vec P-e\,\vec A(\vec R)\,,
\label{20}
\end{eqnarray}
where $M = \sum_a m_a$ and $e = \sum_a e_a$,
and relative coordinates
\begin{eqnarray}
\vec x_a &=& \vec r_a-\vec R\,, \label{21}\\
\vec q_a &=& \vec p_a-\frac{m_a}{M}\,\vec P\,, \label{22}
\end{eqnarray}
such that
\begin{eqnarray}
\bigl[x_a^i\,,\,q_b^j\bigr] &=&
i\,\delta^{ij}\,\biggl(\delta_{ab}-\frac{m_b}{M}\biggr)\,, \label{23}\\
\bigl[R^i\,,\,P^j\bigr] &=& i\,\delta^{ij}\,,  \label{24}\\
\bigl[x_a^i\,,\,P^j\bigr] &=& \bigl[R^i\,,\,q_a^j\bigr] = 0\,. \label{25}
\end{eqnarray}
Next, we perform a canonical transformation $\phi$  
\begin{equation}
H' = e^{-i\,\phi}\,H\,e^{i\,\phi} + \partial_t\phi\,, \label{26}
\end{equation}
which assumes that the characteristic wavelength of 
the electromagnetic field is larger than the size of the system, as follows:
\begin{eqnarray}
\phi &=& \sum_a e_a\,\int_0^1 du\,
\vec x_a\cdot \vec A\bigl(\vec R+u\,\vec x_a\bigr)\nonumber \\ &=&
\sum_a e_a\Bigl[x^i_a\,A^i + \frac{1}{2!}\,x_a^i\,x_a^j\,A^i_{,j} +\ldots
\Bigr]\,.\label{27}
\end{eqnarray}
The scalar potential is transformed to
\begin{equation}
\sum_a e_a\,A^0_a + \partial_t\phi = e\,A^0 -D^i\,E^i - \frac{1}{2!}\,D^{ij}\,E^i_{,j}\,,
\label{28}
\end{equation}
where
\begin{eqnarray}
D^i &=& \sum_a e_a\,x_a^i\,,\\
D^{ij} &=& \sum_a e_a\,x_a^i\,x_a^j\,,
\end{eqnarray}
and $A^0 \equiv A^0(\vec R), \vec E \equiv \vec E(\vec R)$,
similarly $\vec B \equiv \vec B(\vec R)$.
The kinetic momentum is transformed to
\begin{equation}
e^{-i\,\phi}\,\pi_a^j\,e^{i\,\phi} = \tilde \pi_a^j
+\frac{m_a}{M}\,\Pi^j, \label{29}
\end{equation}
where
\begin{equation}
\tilde \pi_a = \vec q_a +\frac{1}{2}\,\Bigl(
e_a\,\vec x_a+\frac{m_a}{M}\,\vec D\Bigr)\times\vec B,
\label{30}
\end{equation}
and the kinetic energy is
\begin{equation}
e^{-i\,\phi}\,\sum_a\frac{\pi_a^2}{2\,m_a}\,e^{i\,\phi} =
\frac{\Pi^2}{2M} 
+ \frac{\vec \Pi}{M}\cdot \vec D\times\vec B
+ \sum_a\frac{\tilde\pi_a^2}{2\,m_a}. 
\label{31}
\end{equation}
Finally, the transformed Hamiltonian takes the form
\begin{equation}
H' =  H_{BP} + H_{\partial E} + H_\Pi, \label{32}
\end{equation}
where
\begin{widetext}
\begin{eqnarray}
H_{BP}  &=& \sum_a\biggl\{\frac{\tilde\pi_a^2}{2\,m_a} 
-\frac{\tilde\pi^4_a}{8\,m_a^3}
-\frac{e_a}{2\,m_a}\,g_a\,\vec s_a\cdot\vec B
+\frac{e_a}{8\,m^3_a}\,\Bigl[
4\,\tilde\pi_a^2 \, \vec s_a\cdot\vec B
+(g_a-2)\,\bigl\{\tilde\pi_a\cdot\vec B\,,\,\tilde\pi_a\cdot\vec s_a\bigr\}\Bigr]
\nonumber\\&&
-\frac{e_a\,(g_a-1)}{2\,m_a^2}\,\vec s_a\times\vec E\cdot\tilde \pi_a\biggr\}
+\sum_{a>b,b} \frac{e_a\,e_b}{4\,\pi}\,\biggl\{\frac{1}{r_{ab}}
-\frac{1}{2\,m_a\,m_b}\,\tilde\pi_a^i\,
\biggl(\frac{\delta^{ij}}{r_{ab}}+\frac{r^i_{ab}\,r^j_{ab}}{r^3_{ab}}
\biggr)\,\tilde\pi_b^j 
\nonumber \\ &&
-\frac{2\,\pi}{3}\,( r_{Ea}^2+r_{Eb}^2) \, \delta^3(r_{ab})
-\frac{2\,\pi\,g_a\,g_b}{3\,m_a\,m_b}\,\vec s_a \cdot\vec s_b\,\delta^3(r_{ab})
+\frac{g_a\,g_b}{4\,m_a\,m_b}\,\frac{s_a^i\,s_b^j}
{r_{ab}^3}\,
\biggl(\delta^{ij}-3\,\frac{r_{ab}^i\,r_{ab}^j}{r_{ab}^2}\biggr)
\nonumber \\ &&
+\frac{1}{2\,r_{ab}^3} \biggl[
\frac{g_a}{m_a\,m_b}\,\vec s_a\cdot\vec r_{ab}\times\tilde\pi_b -
\frac{g_b}{m_a\,m_b}\,\vec s_b\cdot\vec r_{ab}\times\tilde\pi_a +
\frac{(g_b-1)}{m_b^2}\,\vec s_b\cdot\vec r_{ab}\times\tilde\pi_b 
 \nonumber \\ &&
-\frac{(g_a-1)}{m_a^2}\,\vec s_a\cdot\vec r_{ab}\times\tilde\pi_a\bigr)\biggr]\biggr\},
\label{33}
\end{eqnarray}
\begin{eqnarray}
H_{\partial E} &=&-\sum_a\frac{e_a}{6}\,r_{Ea}^2\,\nabla\vec E
-\frac{1}{2!}\,D^{ij}\,E^i_{,j}
-\sum_a\frac{e_a\,(g_a-1)}{4\,m_a^2}\,\epsilon^{ikl}\,\bigl\{s_a^i\,E^k_{,j}\,x_a^j\,,\,
\tilde \pi^l_a\bigr\}, \label{34}
\end{eqnarray}
\begin{eqnarray} 
H_\Pi &=& 
\frac{\vec\Pi^2}{2\,M}+e\,A^0 
- \vec D\cdot\vec E + \vec D\cdot\vec B\times\frac{\vec \Pi}{M} 
- \frac{1}{8\,M^3}\,\Pi^4 
- \frac{1}{2\,M^3}\,\Pi^2\,\vec\Pi\cdot\vec D\times\vec B
\nonumber \\ &&
+ \Pi^i\,Q_{ij}\,\Pi^j + \frac{1}{2}\,\{\Pi^i\,,\,Q_i\}, \label{35}
\end{eqnarray}
\begin{eqnarray}
Q^{ij} &=& -\frac{\delta^{ij}}{2\,M^2}\,\biggl(
\sum_a\frac{\tilde\pi^2_a}{2\,m_a}+\sum_{a>b,b}\frac{e_a\,e_b}{4\,\pi\,r_{ab}}
\biggr)
-\frac{1}{2\,M^2}\,\biggl(
\sum_a\frac{\tilde\pi^i_a\,\tilde\pi^j_a}{m_a}
+\sum_{a>b,b}
\frac{e_a\,e_b}{4\,\pi}\,\frac{r_{ab}^i\,r_{ab}^j}{r^3_{ab}}\biggr)
\nonumber \\ &&
+\sum_a \frac{e_a}{4\,M^2\,m_a}\,\Bigl[
2\,\delta^{ij}\,\vec s_a\cdot\vec B
+(g_a-2)\,B^i\,s_a^j\Bigr], \label{36}
\end{eqnarray}
\begin{eqnarray} 
Q^i  & = &
\sum_a\frac{e_a}{4\,M\,m_a^2}\,\Bigl[
4\,\tilde\pi_a^i\,\vec s_a\cdot\vec B
+(g_a-2)\,B^i\,\tilde\pi_a\cdot\vec s_a
+(g_a-2)\,s_a^i\,\tilde\pi_a\cdot\vec B\Bigr]
\nonumber \\ &&
+\sum_{a>b,b}\frac{e_a\,e_b}{4\,\pi}\,\biggl[
\frac{1}{2\,M\,m_a\,r_{ab}^3}\,(\vec s_a\times\vec r_{ab})^i
-\frac{1}{2\,M\,m_b}\,\biggl(\frac{\delta^{ij}}{r_{ab}}
+\frac{r_{ab}^i\,r_{ab}^j}{r_{ab}^3}\biggr)\tilde\pi_b^j\biggr]
\nonumber \\ &&
-\sum_a\frac{e_a\,(g_a-1)}{2\,m_a\,M}\,(\vec s_a\times\vec E)^i
-\sum_a\frac{1}{4\,M\,m_a^2}\,\{\tilde\pi^2_a\,,\,\tilde\pi_a^i\}. \label{37}
\end{eqnarray}
\end{widetext}
In order to simplify the derivation of the effective Hamiltonian $H_{\rm eff}$ in
Eq. (\ref{15b}), we perform the next canonical transformation $\phi$. 
In this case it should be noted that the $Q^i_0$ operator
\begin{eqnarray}
Q^i_0 &=& Q^i\biggr|_{\vec E=\vec B=0}\nonumber \\
     &=&
\frac{1}{2\,M}
\sum_{a\neq b,b}\frac{e_a\,e_b}{4\,\pi}\,\biggl[
\frac{1}{m_a\,r_{ab}^3}\,(\vec s_a\times\vec r_{ab})^i
\label{38}\\&&
-\frac{1}{\,m_b}\,\biggl(\frac{\delta^{ij}}{r_{ab}}
+\frac{r_{ab}^i\,r_{ab}^j}{r_{ab}^3}\biggr)\,q_b^j\biggr]
-\sum_a\frac{1}{2\,M\,m_a^2}\,\vec q^{\,2}_a\,q_a^i \nonumber
\end{eqnarray} 
can be expressed as a commutator 
\begin{equation}
\vec Q_0 = i\,[H_S\,,\,\vec T], \label{39}
\end{equation}
where
\begin{equation}
\vec T = \frac{1}{2M} \sum_a\left(
\frac{\vec s_a \times \vec q_a}{m_a} - \frac{ q^j_a\,\vec x_a\,q^j_a}{m_a} 
-\sum_{b \neq a} \frac{e_a\,e_b}{4\,\pi} \frac{\vec x_a}{r_{ab}}
\right) \label{40}
\end{equation}
and $H_S$ is the nonrelativistic Hamiltonian of the bound system 
\begin{equation}
H_S = \sum_a\frac{\vec q_a^{\,2}}{2\,m_a} +
\sum_{a>b,b}\frac{e_a\,e_b}{4\,\pi\,r_{ab}}. \label{41}
\end{equation}
Consequently, we assume that
\begin{equation}
\phi = -\vec T\cdot\vec\Pi \label{42}
\end{equation}
and obtain a new Hamiltonian $H''$
\begin{eqnarray}
H'' &=& e^{-i\,\phi}\,H'\,e^{i\,\phi} + \partial_t\phi \label{43}\\
    &=& H' + \delta H, \nonumber
\end{eqnarray}
where
\begin{eqnarray}
\delta H &=& -i\,\biggl[H_S + \frac{\Pi^2}{2\,M} + e\,A^0 - \vec D\cdot\left(
\vec E+\frac{\vec\Pi}{M}\times\vec B\right) 
\nonumber \\ &&
- i\,\partial_t\,,\,\vec T\cdot\vec\Pi\biggr].\label{44}
\end{eqnarray}
We will limit the effective Hamiltonian to terms that are independent,
linear in electromagnetic field strength, and quadratic in electric field, 
thus neglecting the higher order terms in Eq. (\ref{15b}), so
\begin{eqnarray}
H_{\rm eff} &=& \langle\psi_S|H''-H_{BP}\Bigr|_{\vec E = \vec B =0}|\psi_S\rangle 
\label{45}\\&&
+\,\langle\psi_S|H''\,\frac{1}{(\mathcal E_S - H_S)'}\,H''|\psi_S\rangle\nonumber
\end{eqnarray}
The expectation value of $\delta H$ on $\psi_S$ is
\begin{eqnarray}
\langle\delta H\rangle &=& i\,\biggl\langle
\biggl[\vec D\cdot\biggl(\vec E + \frac{\vec\Pi}{M}\times\vec B\biggr)\,,\,
\vec T\cdot\vec\Pi\biggr]\biggr\rangle \nonumber \\ &=&
-\frac{\epsilon^{ijk}}{4\,M}
\biggl\{\biggl(\!\vec E+\frac{\vec\Pi}{M}\times\vec B\!\biggr)^i ,\, \Pi^j\biggr\}
\nonumber \\ &&
\times \sum_a\biggl\langle\biggl(\frac{e_a}{m_a}-\frac{e}{M}\biggr)\,(s_a^k + l_a^k)\biggr\rangle
\nonumber \\ &&
-\frac{1}{6}\langle\vec D\cdot\vec T+ \vec T\cdot\vec D \rangle
\vec\nabla\cdot\vec E 
\label{46}
\end{eqnarray}
The resulting effective Hamiltonian from Eq. (\ref{45}),
after rearrangement, details of which are presented in the following
sections, becomes
\begin{eqnarray}
H_{\rm eff} &=&  
e\,A^0 
+ \frac{\Pi^2}{2\,M}\,\biggl(1-\frac{\mathcal E_S}{M}\biggr) 
-\frac{\Pi^4}{8\,M^3}
-\frac{e}{6}\,R^2\,\nabla\cdot\vec E
\label{47}\\&&
-\frac{e}{2\,M}\,(g+\delta g)\,\vec S\cdot \vec B 
+ \frac{e}{2\,M^2}\,(g-1)\,\vec S\cdot\vec\Pi\times\vec E
\nonumber \\&&
+\frac{e}{2\,M^3}\,\biggl[
\Pi^2\,\vec S\cdot\vec B +
\frac{(g-2)}{2}\,\vec S\cdot\vec \Pi\,\vec B\cdot\vec\Pi\biggr]
-\frac{\alpha_E}{2}\,\vec E^{2}\nonumber
\end{eqnarray}
where $\vec S$ is a global spin operator, and formulas for $R^2$, $g$, $\delta g$, 
and $\alpha_E$ are presented in the following sections.

\subsection{Mean square charge radius}
The mean square charge radius is defined as a coefficient at $\nabla\vec E$,
see Eq. (\ref{47}). It is present in  $H_{\partial E}$
\begin{equation}
H_{\partial E} =
-\sum_a\frac{e_a}{6}\,\nabla\vec E\biggl[ r_{Ea}^2 + x_a^2
+\frac{(g_a-1)}{m_a^2}\,\vec x_a\times\vec q_a\cdot\vec s_a\biggr]
\label{48}
\end{equation}
and also in $\delta H$
\begin{equation}
\langle\delta H\rangle = -\frac{1}{6}\,\nabla \vec E\,
\langle\vec T\cdot\vec D + \vec D\cdot\vec T\rangle. \label{49}
\end{equation}
For the total mean square charge radius $R^2$ of the system, we obtain
\begin{equation}
e\,R^2 = 
\sum_a e_a\,\biggl[
r_{Ea}^2 + \langle x_a^2\rangle
+ \frac{(g_a-1)}{m_a^2}\,\langle\vec x_a\times\vec q_a\cdot\vec s_a\rangle
+ 2\,\langle\vec x_a\cdot\vec T\rangle\biggr]. \label{50}
\end{equation}
The first two terms are widely known, the third one was recently discovered by
Flambaum et al. \cite{flambaum}, and the last term is new.
It would be interesting to calculate them
for deuteron, for which the the charge radius is well known from 
atomic isotope shift measurements \cite{garching}. 
However, in the presence of strong interactions
the formula for $R^2$ may change, and this should be verified using the effective chiral
perturbation theory.

\subsection{Electric dipole polarizability}
The energy shift Eqs. (\ref{45}) due to the electric dipole polarizability Eq. (\ref{47}) is
\begin{equation}
-\frac{\alpha_E}{2}\,\vec E^2 = \langle\psi_S|(\vec D+\delta \vec D)\cdot\vec E\,\frac{1}{\mathcal E_S - H_S}
(\vec D + \delta \vec D)\cdot\vec E|\psi_S\rangle
\end{equation}
so one obtains for $\alpha_E$ 
\begin{equation}
\alpha_E = \frac{2}{3}\,\langle\psi_S|(\vec D+\delta \vec D)\,\frac{1}{H_S - \mathcal E_S}
(\vec D + \delta \vec D)|\psi_S\rangle.
\end{equation}
The relativistic correction to the electric dipole operator
$\delta \vec D$ comes from Eq. (\ref{44})
\begin{equation}
-\delta \vec D\cdot \vec E = -i\,\biggl[e\,A^0 - i\,\partial_t\,,\,\vec T\cdot\vec\Pi\biggr],
\end{equation}
so 
\begin{equation}
\delta\vec D = e\,\vec T,
\end{equation}
and $\vec T$ is defined in Eq. (\ref{40}).
In all the previous calculations of the electric dipole polarizability of nuclei,
the contribution coming from $\delta D$ was missing. 
This correction is particularly important for muonic atoms, 
where nuclear polarizability effects are large.

\subsection{Kinetic energy}
When the electromagnetic field is neglected, the expectation value of
$H''-\mathcal E_S$ is 
\begin{equation}
H_{\rm eff} = \frac{\Pi^2}{2\,M} - \frac{\Pi^4}{8\,M^3} + 
\Pi^i\,\Pi^j\,\langle Q^{ij}_0\rangle, \label{51}
\end{equation}
where
\begin{eqnarray}
 Q^{ij}_0 &=& Q^{ij}\Bigr|_{\vec E = \vec B = 0}\nonumber \\
&=& -\frac{\delta^{ij}}{2\,M^2}\,\biggl(
\sum_a\frac{\vec q^{\,2}_a}{2\,m_a}+\sum_{a>b,b}\frac{e_a\,e_b}{4\,\pi\,r_{ab}}
\biggr)
\nonumber \\ &&
-\frac{1}{2\,M^2}\,\biggl(
\sum_a\frac{q^i_a\,q^j_a}{m_a}
+\sum_{a>b,b}
\frac{e_a\,e_b}{4\,\pi}\,\frac{r_{ab}^i\,r_{ab}^j}{r^3_{ab}}\biggr). \label{52}
\end{eqnarray}
The expectation value of the second term vanishes,
while that of the first term is $\mathcal E_S$, so
\begin{eqnarray}
H_{\rm eff} &=& \frac{\Pi^2}{2\,M}\,\biggl(1-\frac{\mathcal E_S}{M}\biggr) -
\frac{\Pi^4}{8\,M^3}\nonumber \\
 &\approx&  \frac{\Pi^2}{2\,(M+\mathcal E_S)} - \frac{\Pi^4}{8\,(M+\mathcal E_S)^3}.
\label{53}
\end{eqnarray}
$H_{\rm eff}$ is a kinetic energy with the total mass being the sum of individual
masses and the binding energy, as it should be. This is in agreement with
Eq. (\ref{47}).

\subsection{Spin in the external homogenous electric field}
Hereinafter we assume that the electric and the magnetic fields are homogenous. 
The magnetic moment $\mu$ of the compound system is defined as
\begin{equation}
\vec\mu = \biggl\langle \sum_a \frac{e_a}{2\,m_a}\,(\vec l_a + g_a\,\vec s_a)\biggr\rangle\equiv
\frac{e}{2\,M}\,g\,\vec S,  \label{54}
\end{equation}
where the last equation defines the $g$-factor,
and $\vec l_a = \vec x_a\times\vec q_a$ and $\vec S = \sum_a (\vec l_a + \vec s_a)$.
The coupling of the static magnetic moment to the magnetic field is
\begin{equation} 
H_{\rm eff} =-\vec\mu\cdot\vec B\,. \label{55}
\end{equation}
When the system moves, the magnetic moment couples to the electric field as follows
\begin{equation}
H_{\rm eff} = \frac{\vec \Pi\times\vec E}{2\,M}\cdot
\sum_a\left\langle\frac{e_a\,(g_a-1)}{m_a}\,\vec s_a\right\rangle + \delta H,
\label{56}
\end{equation}
where
\begin{equation}
\delta H =
\frac{\vec \Pi\times\vec E}{2\,M}\,\cdot
\sum_a\biggl\langle\biggl(\frac{e_a}{m_a}-\frac{e}{M}\biggr)\,(\vec s_a + \vec l_a)\biggr\rangle.
\label{57}
\end{equation}
After combining both terms
\begin{eqnarray}
H_{\rm eff} &=&\frac{\vec \Pi\times\vec E}{2\,M}\cdot
\sum_a\left\langle \frac{e_a}{m_a}\,(g_a\,\vec s_a + \vec l_a) -
\frac{e}{M}\,(\vec l_a + \vec s_a)\right\rangle
\nonumber \\ &=& \frac{e}{2\,M^2}\,(g-1)\,\vec S\cdot\vec\Pi\times\vec E
\label{58}
\end{eqnarray}
the coupling of the moving spin to the electric field coincides with that in Eq. (\ref{47}).

\subsection{Spin in the external homogenous magnetic field}
Corrections of order $O(\vec\Pi^2)$ to the coupling of the spin to the magnetic field
are
\begin{equation}
H_{\rm eff} = \Pi^i\,\langle Q_B^{ij}\rangle\,\Pi^j + \delta H, \label{59}
\end{equation}
where the part of $Q^{ij}$ that is linear in $\vec B$ is
\begin{eqnarray}
\langle Q_B^{ij} \rangle &=& \frac{1}{2\,M^2}\,\sum_a\frac{e_a}{2\,m_a}\,\Bigl\langle
\delta^{ij}\,\vec x_a\times\vec q_a\cdot\vec B
-(\vec x_a\times\vec B)^i\,q_a^j 
\nonumber \\ &&
- q_a^i\,(\vec x_a\times\vec B)^j
+\Bigl(2\delta^{ij}\,\vec s_a\cdot\vec B
+(g_a-2)B^i\,s_a^j\Bigr)\Bigr\rangle\nonumber \\ &=&
\frac{1}{2\,M^2}\,\sum_a\frac{e_a}{2\,m_a}\bigl\langle
2\,\delta^{ij}\,(\vec l_a+\vec s_a)\cdot\vec B - B^i\,l_a^j 
\nonumber \\&&
+(g_a-2)\,B^i\,s_a^j\bigr\rangle,
\label{60}
\end{eqnarray}
where we used the expectation value identity
\begin{equation}
\langle x_a^i\,q_a^j\rangle = \frac{1}{2}\,\langle x_a^i\,q_a^j - x_a^j\,q_a^i\rangle.
\label{61}
\end{equation}
The contribution from the additional canonical transformation Eq. (\ref{46}) is
\begin{eqnarray}
\delta H &=& 
-\frac{\vec\Pi\times\vec B}{2 M^2}
\sum_a\biggl\langle\!\biggl(\frac{e_a}{m_a}-\frac{e}{M}\biggr)
\vec\Pi\times(\vec s_a + \vec l_a)\biggr\rangle.
\label{62}
\end{eqnarray}
The total $O(\vec\Pi^2)$ interaction takes the form
\begin{equation} 
H_{\rm eff} = \frac{e}{2\,M^3}\,\biggl[
\Pi^2\,\vec S\cdot\vec B +
\frac{(g-2)}{2}\,\vec S\cdot\vec \Pi\,\vec B\cdot\vec\Pi\biggr]
\label{63}
\end{equation}  
and coincides with that in Eq. (\ref{47}).

\section{Magnetic moment}
The relativistic corrections to the magnetic moment of bound states
with arbitrary particle masses have already been considered in the literature 
\cite{hegstrom, magnetic, eides} and very recently in \cite{czarnecki}.
Here we rederive the general formula for the arbitrary state, 
obtain the known result for the magnetic moment 
of hydrogen-like ions in the S state, and confirm and obtain a more 
accurate result for positronium ion Ps$^-$ in the ground state.
Consider the relativistic interaction with the magnetic field resulting
from $H_{BP}$ in Eq. (\ref{33}), and neglect the terms quadratic in $\vec B$.
\begin{widetext}
\begin{eqnarray}
\delta H &=& -\sum_a\frac{e_a}{2\,m_a}\,g_a\,\vec s_a\cdot \vec B+
\sum_a\,\frac{1}{4\,m_a^3}\,\Bigl[
q_a^2\,\vec D_a\times\vec q_a\cdot\vec B
+2\,e_a\,q_a^2\,\vec s_a\cdot\vec B
+e_a\,(g_a-2)\,\vec q_a\cdot\vec s_a\,\vec q_a\cdot\vec B\Bigr]
\nonumber \\ &&
+\sum_{a \neq b,b}\,\frac{e_a\,e_b}{4\,\pi}\biggl[
-\frac{1}{4\,m_a\,m_b}\,q_a^i\biggl(\frac{\delta^{ij}}{r_{ab}}+
\frac{r_{ab}^i\,r_{ab}^j}{r_{ab}^3}\biggr)\,(\vec D_b\times\vec B)^{\,j}
\nonumber \\ &&
+\frac{1}{4\,r_{ab}^3}\,\frac{g_a}{m_a\,m_b}\,
(\vec s_a\times\vec r_{ab})\cdot(\vec D_b\times\vec B)
-\frac{1}{4\,r_{ab}^3}\,\frac{(g_a-1)}{m_a^2}\,
(\vec s_a\times\vec r_{ab})\cdot(\vec D_a\times\vec B)\biggr],
\label{64}
\end{eqnarray}
\end{widetext}
where
\begin{equation}
\vec D_a = e_a\,\vec x_a + \frac{m_a}{M}\,\vec D.
\label{65}
\end{equation}
Eq. (\ref{64}) agrees with the former result of Hegstrom \cite{hegstrom},
and is essentially the same as that in Ref. \cite{magnetic}.
For S-states, only the spin dependent terms in (\ref{64}) contribute, so
\begin{eqnarray}
\delta H 
&=& - \sum_a \frac{e_a}{2 m_a} \vec s_a \cdot \vec B 
\biggl\{ g_a - \frac{\vec q_a^{\,2}}{m_a^2} \left(  \frac{2}{3} 
+ \frac{g_a}{6} \right) 
\nonumber \\ &&
+ \frac{1}{3} \sum_{b \neq a} \frac{e_b}{4 \pi} 
\frac{ \vec r_{ab}}{r_{ab}^3} \cdot \left[ \frac{g_a}{m_b} \vec D_b  
- \frac{g_a - 1}{m_a} \vec D_a \right] \biggr\}. 
\label{66}
\end{eqnarray}
First we will consider the hydrogen-like ion. 
It is a system consisting of one electron of mass $m$ and charge $-e$,  
and the nucleus of charge $Ze$ and mass $m_N$. 
We neglect the spin of the nucleus, so the Hamiltonian is
\begin{eqnarray}
\delta H &=&  \frac{e}{2 m} \vec s \cdot \vec B 
\biggl\{ g_e - \frac{\vec q^{\,2}}{m^2} \biggl(\frac{2}{3} + \frac{g_e}{6}\biggr) 
 + \frac{Z \alpha}{3\,M\,r_{eN}} 
\nonumber\\&&
\times\biggl[g_e\,\biggl(-Z\frac{M}{m_N} + (Z - 1)\frac{m_N}{M}\biggr)
\nonumber \\ &&   
+ (g_e - 1)\,\biggl(\frac{M}{m} + (Z-1)\frac{m}{M}\biggr) \biggr] \biggr\}.  
\label{67}
\end{eqnarray}
The correction $\delta g$ is given by the expectation value of the Hamiltonian on 
the hydrogen-like system state $\phi =  \phi_{nlm_l}$, 
where $n, l, m_l$ are respective quantum numbers
\begin{equation}
\langle \delta H \rangle 
= \frac{e}{2 m} \vec s \cdot \vec B (g_e + \delta g),  
\label{68}
\end{equation}
where 
\begin{eqnarray}
\delta g &=& -\frac{(Z\,\alpha)^2}{3\,n^2\,(1+x)^2}\,\biggl[
-\frac{g_e}{2} + 4 - \frac{1}{1+x}
\nonumber \\ &&
+Z\,x^2\,\biggl(g_e + \frac{1}{1+x}\biggr)\biggr]
\label{69}
\end{eqnarray}
and $x= m/ m_N$, in agreement with Refs. \cite{magnetic, eides}. 

In the case of the positronium ion Ps$^-$ in the ground state, 
the spin comes only from the positron, since two electrons 
are in the singlet state. 
Hence from the beginning we neglect terms proportional to electron spin,
and assume that indices $1, 2$ refer to electrons, while index $3$ refers to the positron. 
\begin{eqnarray}
\delta H
&=& - \frac{e}{2\,m} \vec s_3 \cdot \vec B 
\left\{ g_e - \frac{\vec q_3^{\;2}}{m^2} \left(  \frac{2}{3} + \frac{g_e}{6} \right) \right. \label{70} \\
&&  - \frac{\alpha}{9\,m} 
 \left[ \frac{5}{3} \left( \frac{1}{r_{13}} + 
\frac{\vec r_{13} \cdot \vec r_{23}}{r_{13}^3} \right) 
+ g_e \frac{\vec r_{13} \cdot (\vec r_{12} - \vec r_{23})}{r_{13}^3} \right. \nonumber \\
&& \left. \left. + \frac{5}{3} \left( \frac{1}{r_{23}} 
+ \frac{\vec r_{13} \cdot \vec r_{23}}{r_{23}^3} \right) 
- g_e \frac{\vec r_{23} \cdot (\vec r_{12} + \vec r_{13})}{r_{23}^3} \right] \right\} 
\nonumber
\end{eqnarray}
As previously, we find a correction to the magnetic moment of Ps$^-$ 
from the expectation value of the Hamiltonian (\ref{70}).  
\begin{equation}
\langle \delta H\rangle = - \frac{e}{2 m} \vec s_3 \cdot \vec B\, g_{{\rm Ps}^-}. \label{71}\\
\end{equation}
Of note, $g_{{\rm Ps}^-}$ is defined differently 
from the g-factor in Eq. (\ref{54}). Following this definition and Eq. (\ref{70})
$g_{{\rm Ps}^-}$ is
\begin{eqnarray}
g_{{\rm Ps}^-} &=& g_e + \alpha^2\,\delta g, \label{72} \\
\delta g &=& 
-\biggl\langle \vec p_3^{\,2} \left( \frac{2}{3} + \frac{g_e}{6} \right)  
+ \frac{1}{r_{13}}\left(\frac{10}{27} + \frac{2}{9}\,g_e\right)  
\nonumber \\ &&
+ \frac{\vec r_{13} \cdot \vec r_{23}}{r_{13}^3}\left(\frac{10}{27} - \frac{4}{9}\,g_e\right) \biggr\rangle,
\label{73}
\end{eqnarray}
where for convenience we used in the last equation atomic units,
so above matrix elements are dimensionless.
Table I presents expectation value of operators 
in the ground state of Ps$^-$  calculated numerically, 
\begin{table}[ht]
\caption{Expectation values of operators in (\ref{73}) on the ground state of
  Ps$^-$ in atomic units, fundamental constants are from Ref. \cite{codata}}
\begin{tabular}{l@{\hspace{0.5cm}}w{2.14}}
\hline
\hline
Energy &  -0.262\,005\,070 \\
$\langle \vec p_3^{\;2} \rangle$ & 0.257\,532\,962 \\
$\langle \frac{1}{r_{13}}\rangle$ & 0.339\,821\,023 \\
$\langle \frac{\vec r_{13} \cdot \vec r_{23}}{r_{13}^{\;3}}\rangle$ & 0.046\,478\,421 \\
$\delta g$    & -0.510\,551\,028(1)\\
$g_e$         &  2.002\,319\,304\\
$g_{{\rm Ps}^-}$ & 2.002\,292\,117(3) \\
\hline
\hline
\end{tabular}
\end{table}
where the uncertainty for our total g-factor $g_{{\rm Ps}^-}$  
is estimated by $ 2\,\alpha^4 \,\delta g$.
Our value for $\delta g$ is in agreement with the one obtained in
\cite{czarnecki} $-0.51(1)$, but is significantly more accurate and includes
the leading QED effects. Surprisingly we do not agree with the corresponding formula obtained 
in \cite{czarnecki}, which is much different and not equivalent to that of
ours in Eq. (\ref{73}), and also we disagree with their total g-factor $g_{{\rm Ps}^-} = 2.004\,61(1)$.

\section{Summary}
We have presented an approach to derive an effective Hamiltonian that governs the dynamics
of the whole bound system from individual Hamiltonians of its ingredients, including 
leading relativistic corrections. 
This approach is based on our two former works \cite{compound, magnetic}, 
and in comparison to them it is much simpler.
We derived a formula for the charge radius, which can be used for systems such as nuclei. 
Besides the known terms, it includes new terms which until now have not been taken into account
in the calculation of nuclear charge radii. Similarly, the electric dipole
polarizability includes corrections to the electric dipole moment  $\delta D$ that have been
omitted in all previous calculations of the nuclear polarizability.
The obtained formula for the magnetic moment is in agreement with that obtained previously
\cite{hegstrom, magnetic, eides}
and we improve the result published recently for the positronium ion \cite{czarnecki}. 
The presented approach can also be used for nuclei, to calculate their electromagnetic moment,
but this requires incorporation of strong interactions via the chiral perturbation theory.
It is especially important in view of very accurate results for nuclear charge radii
differences between isotopes obtained from atomic spectroscopy \cite{garching, halo}.
    
\section*{Acknowledgments}
Authors would like to acknowledge support by NCN grant 2012/04/A/ST2/00105.

\end{document}